\documentclass[preprint,12pt]{elsarticle}

\usepackage{amssymb}
\usepackage{amsmath}
\usepackage{booktabs}
\usepackage{microtype}
\usepackage{array}
\usepackage{graphicx}
\usepackage{xcolor}
\usepackage{float}
\usepackage[hidelinks]{hyperref}

\journal{Chaos, Solitons \& Fractals}

\newcommand{\lamhat}{\widehat{\lambda}}

\newcommand{\fexact}{F_{\mathrm{exact},K}}
\definecolor{revgreen}{RGB}{0,125,45}
\newcommand{\rev}[1]{#1}
\newenvironment{revision}{}{}

\begin{document}

\begin{frontmatter}

\title{When More Data Become Less Informative: Finite-Precision Periodicization and Collapse of Forecast-Error Lyapunov Estimates}

\author[aff1]{Andrei Velichko\corref{cor1}\fnref{orcid1}}
\ead{velichkogf@gmail.com}
\author[aff2]{Viet-Thanh Pham\fnref{orcid2}}

\cortext[cor1]{Corresponding author.}
\fntext[orcid1]{ORCID: 0000-0002-9341-1831.}
\fntext[orcid2]{ORCID: 0000-0001-5151-9812.}

\affiliation[aff1]{organization={Institute of Physics and Technology, Petrozavodsk State University},
            addressline={33 Lenina Prospekt},
            city={Petrozavodsk},
            state={Republic of Karelia},
            postcode={185910},
            country={Russia}}

\affiliation[aff2]{organization={Faculty of Electronics Technology, Industrial University of Ho Chi Minh City},
            city={Ho Chi Minh City},
            country={Vietnam}}

\begin{abstract}
\begin{revision}
Equation-free forecast-error analysis is useful when governing equations, Jacobians, tangent dynamics, or full state vectors are unavailable and only scalar time series are observed. In this setting, the multi-horizon forecast-error curve can provide both an instability estimate and nonlinear predictive features, making its numerical reliability important. We show that long finite-precision chaotic records can enter an information-saturation regime in which increasing record length $N$ (the total number of analyzed samples) adds repeated histories rather than new dynamical information. For the \texttt{float32} logistic map, exact recurrence progressively fills the training catalogue with duplicate histories and identical futures, suppressing forecast-error growth and driving the estimated Lyapunov slope toward zero. Across 10,000 initial conditions, exact recurrence occurred by iteration 7612; across 28 stratified seeds, collapse onset strongly followed the recurrence-occupancy scale. A two-dimensional H\'enon-map control confirms that recurrence and estimator failure are separate events: the \texttt{float32} trajectory enters an exact machine-state cycle after a transient of $\mu=103{,}039$ samples, with cycle period $P=12{,}268$, yet the forecast-error estimate remains close to the positive Jacobian/QR reference until catalogue saturation near $N=200{,}000$. Casting an already generated \texttt{float32} record to \texttt{float64} does not restore lost information, whereas genuine \texttt{float64} generation, restarted short trajectories, and one-ULP perturbations prevent collapse over the tested range. A recursive local-linear predictor exhibits the same endpoint at a later, predictor-dependent threshold. The results identify a recurrence-to-catalogue-saturation failure mode specific to forecast-error/analogue inference rather than Lyapunov estimation in general.
\end{revision}
\end{abstract}


\end{frontmatter}

\noindent{\small\textbf{Publication information.} This is the authors' accepted-manuscript version, not the publisher-formatted Version of Record. Published as: A. Velichko and V.-T. Pham, \emph{Chaos, Solitons \& Fractals} \textbf{213} (2026), 119363. \url{https://doi.org/10.1016/j.chaos.2026.119363}.}\par\medskip

\section{Introduction}
\label{sec:intro}

A positive largest Lyapunov exponent (LLE) is one of the standard quantitative signatures of chaos. Classical time-series methods infer exponential instability from the evolution of nearby reconstructed trajectories or tangent-space surrogates \cite{Wolf1985,Rosenstein1993,Parlitz2016}. More recently, data-driven and machine-learning formulations have broadened this viewpoint: reservoir and supervised-learning models can recover chaotic instability information from observed trajectories \cite{Pathak2017,Ayers2022}, while our previous work estimated the LLE directly from the logarithmic growth of out-of-sample multi-horizon forecast error \cite{Velichko2025}. In that formulation, a predictor is trained on a scalar record, absolute forecast errors are geometrically averaged, and the slope of $\log E(h)$ versus horizon $h$ serves as an LLE proxy. The method is attractive for short experimental records because it does not require explicit governing equations.

\begin{revision}
The equation-free setting is a complementary use case rather than a substitute for model-based Lyapunov analysis. Jacobian/tangent-space methods are natural when the governing equations or a differentiable dynamical model are available, and the LBE/interval-extension approach likewise exploits alternative arithmetic realizations of known equations \cite{Mendes2016}. Forecast-error profiling addresses the opposite situation in which only an observed scalar record may be accessible and the governing equations, tangent dynamics, and full state vector are unknown. Recent extensions of this forecast-error philosophy illustrate why reliability of the profile itself matters. FEG-Pro treats the multi-horizon error curve as a structured object that yields not only an instability-related slope but also local slopes, curvature, residual roughness, monotonicity, fit diagnostics, and forecast-error distribution entropy as candidate nonlinear features \cite{Velichko2026FEGPro}. A related equation-free formulation estimates negative contraction rates from ensembles of short scalar trajectories without an analytical Jacobian \cite{Velichko2026Negative}. Multi-horizon kNN profile geometry has also been used as a screening layer for Mittag--Leffler-compatible dynamics \cite{NGbo2026Mittag}, while forecast-error-growth descriptors have already been evaluated as exploratory nonlinear predictability features in a peer-reviewed respiratory-signal application \cite{VelichkoHuyut2027}. These developments motivate the broader reliability question considered here: when finite-precision recurrence changes the forecast-error profile itself, the consequence can extend beyond a single LLE-like number to other descriptors derived from the same predictive geometry.
\end{revision}

Finite precision introduces a separate and well-established issue. A deterministic digital map has a finite state space; once an exact machine state repeats, its complete future must also repeat. Consequently, sufficiently long finite-precision trajectories consist of a transient followed by an exact digital cycle. Periodicity induced by floating-point representation has been documented for the logistic map and other low-dimensional maps \cite{Persohn2012,Galias2021}, and broader studies have characterized how precision, rounding, switching, and state-space structure influence digital-chaos complexity and degradation \cite{Antonelli2018,Kloewer2023,Fan2021}.

\begin{revision}
Several closely related finite-precision results precede the present study and define a useful hierarchy of questions. Mendes and Nepomuceno \cite{Mendes2016} used lower-bound error (LBE) between arithmetic interval extensions to estimate a positive LLE and demonstrated the approach for the Logistic and H\'enon maps and the Lorenz, R\"ossler, and Mackey--Glass systems. Nepomuceno et al. \cite{Nepomuceno2017} then used LBE to show that pointwise numerical reliability of Logistic- and H\'enon-map pseudo-orbits can be lost in fewer than $10^2$ double-precision iterations under their criterion. The explicit critical-simulation-time relation appears in Nepomuceno et al. \cite{Nepomuceno2018}. Writing $p=\log_{10}u<0$ for the logarithmic roundoff scale, where $u$ is the unit roundoff ($u=2^{-24}$ for IEEE \texttt{float32} and $u=2^{-53}$ for IEEE \texttt{float64} under round-to-nearest), and $\lambda_{10}$ for an LLE expressed in base-10 logarithmic units, the relation can be written as
\begin{equation}
 T_s=\frac{\log_{10}(D/2)-p}{\lambda_{10}},
 \label{eq:Ts}
\end{equation}
where $D$ is the phase-space diameter. This is a trajectory-fidelity scale: it concerns when round-off-amplified pseudo-orbit separation becomes comparable with the attractor scale, not when a finite-state trajectory first repeats exactly. Exact machine-state recurrence is a different, later graph-theoretic event. Generic random-mapping theory likewise predicts collision/recurrence scales of order the square root of the accessible state count \cite{Flajolet1990}. The present work concerns a third scale: the still later record length at which repeated recurrence has populated a forecast catalogue with enough duplicate histories and identical futures to alter a forecast-error LLE estimate. Schematically, for the reduced-precision cases studied here,
\begin{equation}
 T_s \ll T_{\mathrm{recurrence}} \lesssim T_{\mathrm{FE}}^{\mathrm{collapse}},
 \label{eq:timescales}
\end{equation}
with the important qualification that the last scale is estimator- and acquisition-dependent rather than a universal property of the dynamical system.
\end{revision}

A particularly important implication of this literature is that reduced precision should not be viewed only as a small numerical perturbation. Digital-chaos studies describe finite-word-length implementations as effective dynamical systems with their own reachable states, transients, cycle structure, and basin organization \cite{Fan2021,Galias2021,Kloewer2023}. Several mitigation schemes have therefore been developed to extend digital periods or break repeated-state trapping, including self-perturbation and delay/disturbance strategies \cite{Merah2021,Li2024}. The engineering motivation is substantial: chaotic maps are implemented on FPGA, DSP, microcontrollers, and embedded processors for pseudorandom generation, cryptography, control, and signal processing \cite{Wang2016,DeLaFraga2017,FloresVergara2019,Murillo2022,Huynh2019}. In such settings, single precision or fixed-point arithmetic can be a deliberate trade-off for speed, memory, area, or power.

\begin{revision}
The contribution of the present study is therefore not the discovery of finite-precision periodicity, not a new general cure for digital periodicity, and not a claim that LLE estimation as a whole collapses on recurrent data. We isolate a specific long-record failure regime of \emph{forecast-error/analogue} Lyapunov inference: exact digital recurrence progressively fills the predictive catalogue with duplicate histories and identical futures, so increasing nominal sample count can reduce effective dynamical information. For the local forecast architectures studied here, this produces a measurable transition from a positive forecast-error growth slope to an artificial near-zero slope once the predictor's required neighborhood multiplicity is saturated. Tangent-space/Jacobian methods, classical divergence estimators, and LBE/interval-extension methods do not use this catalogue mechanism and are outside this collapse claim. Indeed, recent experimental work has successfully extended LBE-based LLE estimation to physically implemented Jerk and Chua circuits \cite{Nazare2026}, further emphasizing that the present failure mode is methodological rather than a universal loss of Lyapunov information.
\end{revision}

The mechanism is simple but consequential. Suppose that after a transient $\mu$ a digital orbit has period $P$. Then
\begin{equation}
 \mathbf{v}_{n+P}=\mathbf{v}_{n}, \qquad x_{n+P+h}=x_{n+h},
\end{equation}
for any history vector $\mathbf{v}_n$ and forecast horizon $h$ fully contained in the cycle. If the training set contains $K$ exact copies of a test history, a $K$-nearest-neighbor predictor can return an identical future, so $e(h)=0$ up to the numerical floor. A forecast-error LLE estimator then observes a nearly flat $\log E(h)$ profile and reports $\widehat{\lambda}\approx0$, even though the intended real-valued system may still have a positive LLE. This is different from the known use of precision- or rounding-induced pseudo-orbit separation to estimate positive Lyapunov growth \cite{Peixoto2018,Zhou2019}: our focus is the opposite long-record limit, where finite-state recurrence suppresses forecast-error growth.

This question also changes the usual interpretation of ``more data.'' For ordinary statistical learning, larger training sets generally improve neighborhood coverage. In a deterministic finite-state chaotic record, however, sufficiently late samples may add exact copies rather than new dynamical information. The practical issue is therefore not whether a reduced-precision processor computes its arithmetic incorrectly. On the contrary, it may compute the specified finite-precision map exactly, while the long-time dynamics of that map has become periodic. For on-device chaos diagnostics, edge computing, embedded experiments, or digitally generated test signals, this distinction can determine whether a long record still represents the instability of the intended real-valued system.

\rev{This motivates an explicit distinction between nominal sample size and effective information content: a data set may continue to grow while the number of dynamically distinct predictive histories has already saturated.}

The present study isolates this effect using the logistic map at $r=4$, for which the real-valued reference is $\lambda=\ln2$. First, an ESP32 implementation is validated against NumPy \texttt{float32}; the trajectories agree bit-for-bit over $1{,}000{,}001$ stored values, so large systematic sweeps can subsequently be carried out in software without treating hardware and Python \texttt{float32} as different dynamical objects. Second, 10,000 random initial conditions are used to characterize exact digital recurrence and the dominant period structure. Third, we measure how the kNN forecast-error LLE changes as one continuous record is lengthened and directly compare the transition with exact train--test neighbor saturation. Fourth, the same total data budget is redistributed among independently restarted short trajectories to test whether acquisition strategy can preserve access to the pre-periodic chaotic transient. Finally, a stratified multi-seed experiment shows that the collapse is not specific to $x_0=0.2$ and that its onset is quantitatively linked to the transient $\mu$, period $P$, neighbor count $K$, and train fraction.

\begin{revision}
Five complementary controls isolate the mechanism. We quantify saturation of distinct predictive histories as $N$ grows; repeat the kNN analysis after casting an already generated \texttt{float32} record to \texttt{float64} without regenerating it; and apply an exactly one-ULP \texttt{float32} perturbation every 100 iterations as a causal control for fixed-cycle recurrence. We also replace local-constant kNN averaging by a recursively iterated local-affine predictor to test architecture dependence. Finally, the recurrence--saturation experiment is repeated for the two-dimensional H\'enon map at $a=1.4$, $b=0.3$, with an independent Jacobian/QR Lyapunov reference while forecast-error inference still uses only the scalar $x_n$ history. Together, these controls distinguish information lost during generation from precision used during subsequent analysis, separate the data-level saturation mechanism from predictor-dependent details of its onset, and test whether the mechanism extends beyond the special logistic $r=4$ implementation.
\end{revision}

The remainder of this paper is organized as follows. Section~\ref{sec:methods} describes the finite-precision logistic-map realizations, hardware validation, forecast-error LLE estimator, and long-record and restarted acquisition protocols. Section~\ref{sec:results} presents the digital-cycle statistics, the long-record LLE collapse, its relation to exact-neighbor saturation, and the multi-seed recurrence-scale analysis. \rev{The Methods and Results additionally test predictor dependence with a recursive local-linear control and dynamical generality with the H\'enon map and an independent Jacobian/QR Lyapunov reference.} Section~\ref{sec:discussion} discusses the mechanism, practical implications for reduced-precision computation, and the broader interpretation of forecast-error saturation as a recurrence diagnostic. Finally, Section~\ref{sec:conclusions} summarizes the main conclusions.

\section{Methods}
\label{sec:methods}

\subsection{Reference system and numerical realizations}

We consider the logistic map
\begin{equation}
 x_{n+1}=r x_n(1-x_n), \qquad r=4.
 \label{eq:logistic}
\end{equation}
For the ideal real-valued system at $r=4$, the reference LLE is
\begin{equation}
 \lambda_{\mathrm{ref}}=\ln 2 \approx 0.693147.
 \label{eq:lref}
\end{equation}

Two numerical realizations were used. The reduced-precision realization evaluates Eq.~\eqref{eq:logistic} in IEEE-like single precision using the operation order
\begin{equation}
 x_{n+1}=\mathrm{float32}\left(4.0_f\,x_n\,(1.0_f-x_n)\right),
 \label{eq:f32}
\end{equation}
where all operands are single-precision values. A \texttt{float64} realization from the same quantized initial condition serves as a higher-precision numerical control.

The hardware validation used an ESP32 implementation and a long record with the initial condition stored as $x_0=0.20000000298023224$. The record contained $1{,}000{,}001$ values. The corresponding NumPy \texttt{float32} trajectory was generated with the same operation order. After establishing exact agreement, the systematic initial-condition and data-budget sweeps were performed in Python.

\subsection{Exact finite-precision cycles}

For a deterministic finite-precision trajectory, we define the first exact recurrence by
\begin{equation}
 x_{\mu+P}=x_{\mu},
 \end{equation}
with equality tested at the machine-bit level. Here $\mu$ is the transient length before the digital cycle and $P$ is the exact cycle period. Once this equality occurs, determinism implies
\begin{equation}
 x_{\mu+P+h}=x_{\mu+h}, \qquad h\geq 0.
 \label{eq:repeat}
\end{equation}
Thus, the number of distinct states visited before the first repeated state is $\mu+P$.

Cycle statistics were first evaluated for 10,000 unique random \texttt{float32} initial conditions in $(0,1)$. Exact periods and transient lengths were identified using a vectorized Floyd cycle-detection procedure and verified from the generated trajectories. A set of 130 valid long ESP32 records was retained as an independent hardware spot check of the dominant digital periods.

\subsection{kNN multi-horizon forecast-error LLE}

The LLE estimator follows the forecast-error principle introduced in \cite{Velichko2025}. A history vector of length $Y$ is constructed as
\begin{equation}
 \mathbf{v}_n=(x_n,x_{n+1},\ldots,x_{n+Y-1}).
\end{equation}
For each test history, the $K$ nearest training histories are found in Euclidean distance. The future at horizon $h$ is predicted by averaging the corresponding $K$ training futures. With $x_i(h)$ denoting the true test future and $\widehat{x}_i(h)$ the prediction, the absolute forecast error is
\begin{equation}
 e_i(h)=|x_i(h)-\widehat{x}_i(h)|.
\end{equation}
The geometric mean absolute error (GMAE) is evaluated as
\begin{equation}
 E(h)=\exp\left[\frac{1}{N_t}\sum_{i=1}^{N_t}\ln\left(\max(e_i(h),\varepsilon)\right)\right],
 \label{eq:gmae}
\end{equation}
where $\varepsilon=10^{-12}$ is a numerical floor. The estimated exponent is the least-squares slope
\begin{equation}
 \ln E(h)\approx C+\lamhat h.
 \label{eq:slope}
\end{equation}

The present experiments use $Y=5$, $K=3$, and horizons $h=1,\ldots,5$. These settings deliberately stay close to the previously published positive-LLE formulation so that the present study tests a change in the data-generating regime rather than a new estimator design.

\subsection{Long-record and restarted-trajectory protocols}
\label{sec:protocols}

\begin{revision}
To separate the effect of total data volume from the effect of how the data are acquired, we compare two ways of collecting the same nominal number of samples: one long continuous trajectory and an ensemble of many independently restarted short trajectories. In the long-record protocol, a single finite-precision realization is allowed to evolve continuously and can therefore spend most of a large data budget traversing the same eventual digital cycle. In the restarted protocol, each short realization begins from a distinct initial state and samples its own pre-periodic transient before cycle capture can dominate that segment.
\end{revision}

For the \emph{long-record} protocol, one continuous trajectory of total length $N$ is split chronologically: the first $70\%$ forms the training region and the last $30\%$ the test region. Windows are fully contained within their respective regions, and no random train--test mixing is used.

For the \emph{restarted} protocol, the same total data budget is distributed among independent trajectories of length 100. The trajectories are initialized from distinct \texttt{float32} seeds in $(0,1)$; matched \texttt{float64} trajectories use the same quantized initial states. Whole trajectories, rather than individual windows, are assigned to training and test sets using the same 70/30 fraction. This prevents overlapping windows from the same short trajectory from appearing on both sides of the split. The analysis excludes the first 20 points of each trajectory when forming histories. The total budgets ranged from $10^3$ to $10^6$ stored points.

To diagnose the mechanism of collapse, we additionally record the fraction of test histories whose $K$th nearest-neighbor distance is exactly zero,
\begin{equation}
 \fexact = \frac{\#\{i:d_{i,K}=0\}}{N_t}.
 \label{eq:fexact}
\end{equation}
For $K=3$, $\fexact=1$ means that every queried test history has at least three exact copies in the training set. Because all exact copies of a deterministic cycle have identical futures, the kNN forecast becomes exact apart from the numerical floor in Eq.~\eqref{eq:gmae}.

\subsection{\textcolor{black}{Information saturation and mechanistic controls}}
\begin{revision}
Three complementary analyses were performed using the same ESP32-validated control record and the same $Y=5$, $K=3$, 70/30 chronological split, forecast horizons, and GMAE estimator described above.

First, effective information content was quantified at both the state and history levels. With $Y=5$, each predictor input is a five-sample history, for example $(x_{n-4},x_{n-3},x_{n-2},x_{n-1},x_n)$. Before exact cycle capture, almost every such history is new. Once a finite-precision trajectory enters an exact digital cycle, the same five-state sequences reappear on every traversal of that cycle. The nominal record length can therefore continue to increase even after the number of genuinely distinct predictor histories has nearly stopped growing.

For each nominal record length $N$, machine states were compared bitwise and predictor histories were compared bytewise with no numerical tolerance. Let $U_H(N)$ denote the number of distinct admissible predictor histories and $N_H(N)$ the total number of histories entering the long-record analysis. We report the unique-history fraction
\begin{equation}
 R_H(N)=\frac{U_H(N)}{N_H(N)}.
 \label{eq:uniquehist}
\end{equation}
A decreasing $R_H$ therefore means that additional stored samples predominantly reproduce histories already present in the catalogue. In parallel, $F_{\mathrm{exact},K}$ from Eq.~\eqref{eq:fexact} measures the stronger condition relevant to the $K$-neighbor predictor: a test history must have at least $K$ exact training copies.

Second, an offline high-precision-analysis control was constructed without changing the generated trajectory. The ESP32/NumPy \texttt{float32} sequence was first generated and stored exactly as in the baseline experiment and was then cast elementwise to \texttt{float64} before history construction, distance calculations, neighbor search, prediction, and slope fitting. This operation increases the arithmetic precision of the \emph{analysis} but cannot create new trajectory states or restore bits that were discarded during \texttt{float32} generation. A separately generated \texttt{float64} trajectory from the same quantized initial condition was retained as the generation-precision control.

Third, we implemented a minimal scheduled perturbation as a direct control for the suggestion that occasional perturbations can prevent digital-cycle trapping. Starting from the same $x_0$, the ordinary \texttt{float32} logistic update was applied at every iteration. After every 100th update only, the resulting state was shifted to the immediately adjacent representable \texttt{float32} number using \texttt{nextafter}; the direction toward 0 or 1 was selected pseudorandomly with the fixed seed 20260919. Thus the perturbation magnitude was exactly one unit in the last place (one ULP) at the current state, rather than an additive Gaussian or fixed-amplitude noise. No perturbation was applied during the intervening 99 iterations. Because this scheduled external intervention makes the update non-autonomous, a repeated scalar value in this control is treated as a collision and not as proof of entry into an autonomous periodic orbit. This perturbation arm is used only as a mechanistic control and is not proposed as a new recurrence-mitigation method.
\end{revision}

\subsection{\textcolor{black}{Recursive local-linear predictor control}}
\begin{revision}
To test whether the long-record collapse is specific to the local-constant $K=3$ kNN predictor, we performed a predictor-architecture control using a recursive local-affine model. The input histories, chronological 70/30 split, long and restarted protocols, and forecast horizons $h=1,\ldots,5$ were kept unchanged. For each query history $\mathbf{q}$, the $K_{\mathrm{LL}}=20$ nearest training histories $\mathbf{v}_j$ were used to fit only the one-step transition,
\begin{equation}
 x_{j,+1}=a+\mathbf{b}^{\mathsf T}(\mathbf{v}_j-\mathbf{q}).
 \label{eq:locallinear}
\end{equation}
The fit is centered at the query, so the one-step forecast at $\mathbf{q}$ is the intercept $a$. The intercept is unregularized, while the slope coefficients use a small local-scale ridge term
\begin{equation}
 \alpha=\max\!\left(10^{-12},10^{-8}\,\frac{\operatorname{tr}(D^{\mathsf T}D)}{Y}\right),
 \label{eq:llridge}
\end{equation}
where the rows of $D$ are the centered neighbor vectors. After each one-step prediction, the delay history is advanced with the predicted value, neighbors are recomputed around the new predicted history, and the same one-step local-affine construction is iterated recursively to horizons 1--5. This avoids fitting five unrelated direct $h$-step regressions and provides a local dynamical forecasting architecture distinct from the original kNN average.

The control was evaluated for $N=6000$ to $10^6$ using long \texttt{float32}, long \texttt{float64}, restarted \texttt{float32}, and restarted \texttt{float64} data. For large test sets, at most 8000 uniformly spaced test histories were queried in each condition, with the same subset used for the predictor and exact-neighbor diagnostics. In addition to the forecast-error slope, we recorded $F_{\mathrm{exact},20}$, the fraction of queries whose 20th nearest training-neighbor distance is exactly zero. Because the recursive local-affine model is used as a robustness control rather than independently calibrated as a second high-accuracy LLE estimator, its absolute slope values are not interpreted as unbiased estimates of $\ln2$ at every data budget. The relevant test is whether finite-state catalogue saturation suppresses forecast-error growth selectively in the long \texttt{float32} record while precision and restarted controls remain non-collapsed.
\end{revision}

\subsection{Generality across initial conditions}

To determine whether the collapse is specific to the control seed $x_0=0.2$, 10,000 candidate \texttt{float32} seeds were classified by $(\mu,P)$. A stratified subset of 28 representative initial conditions was then selected across the observed digital periods $P\in\{1,136,143,436,836,4344\}$ and across a broad range of transient lengths. For each seed, long-record \texttt{float32} and matched \texttt{float64} LLE curves were computed over $N=2{,}000$--$100{,}000$.

For a cycle of period $P$, a simple estimate of the data scale at which $K$ exact copies can become available in the training region is
\begin{equation}
 N_{K}^{\mathrm{pred}}\approx \frac{\mu+KP}{f_{\mathrm{train}}},
 \label{eq:predscale}
\end{equation}
with $f_{\mathrm{train}}=0.7$. This is not used as a fitted model; it is an interpretable recurrence scale against which the observed collapse is compared.

\subsection{\textcolor{black}{H\'enon-map generalization control}}
\begin{revision}
To test whether the recurrence-induced collapse is specific to the one-dimensional logistic map at $r=4$, we repeated the analysis for the standard H\'enon map
\begin{align}
 x_{n+1} &= 1-a x_n^2+y_n,\\
 y_{n+1} &= b x_n,
 \label{eq:henon}
\end{align}
with $a=1.4$, $b=0.3$, and $x_0=y_0=0$. This parameter set is a canonical chaotic H\'enon regime used in previous Lyapunov-exponent and finite-precision reliability studies \cite{Mendes2016,Nepomuceno2017}. Exact \texttt{float32} recurrence was defined by bitwise equality of the complete state pair $(x_n,y_n)$; equality of $x_n$ alone was not sufficient. The transient $\mu$ and cycle period $P$ were obtained with Brent's constant-memory cycle algorithm.

An independent reference LLE was calculated from the full two-dimensional dynamics in \texttt{float64}. At each step the tangent basis was propagated by the Jacobian
\begin{equation}
 J_n=
 \begin{pmatrix}
 -2a x_n & 1\\
 b & 0
 \end{pmatrix},
 \label{eq:henonjac}
\end{equation}
followed by QR re-orthogonalization. After a burn-in of 20,000 iterations, logarithmic diagonal growth factors were accumulated for 300,000 steps. The identity $\lambda_1+\lambda_2=\ln|b|$, which follows from $|\det J_n|=|b|$, was used as an internal numerical consistency check.

The forecast-error analysis itself was deliberately left unchanged. Only the scalar $x_n$ record was supplied to the same $Y=5$, $K=3$, horizons 1--5 kNN estimator and the same chronological 70/30 long-record split. Matched long \texttt{float64} and independently restarted length-100 \texttt{float32} trajectories were evaluated as controls, with the first 20 points of each restarted trajectory excluded from history construction exactly as in the logistic-map experiment. For large test sets, at most 12,000 uniformly spaced test histories were queried. At each data budget we recorded the fraction of unique bitwise $(x,y)$ states, the fraction of unique scalar $Y=5$ histories, and $F_{\mathrm{exact},3}$. The recurrence-scale prediction in Eq.~\eqref{eq:predscale} was evaluated without refitting any parameter.
\end{revision}

\section{Results}
\label{sec:results}

\subsection{Hardware validation reveals two distinct finite-precision time scales}

The first result separates two effects that occur on very different time scales. For the control seed, the ESP32 single-precision trajectory and NumPy \texttt{float32} trajectory were identical at the bit level for all $1{,}000{,}001$ stored values. Hence the software implementation can be treated as an exact emulator of the tested hardware arithmetic for subsequent sweeps.

\begin{figure}[t]
\centering
\includegraphics[width=0.98\linewidth]{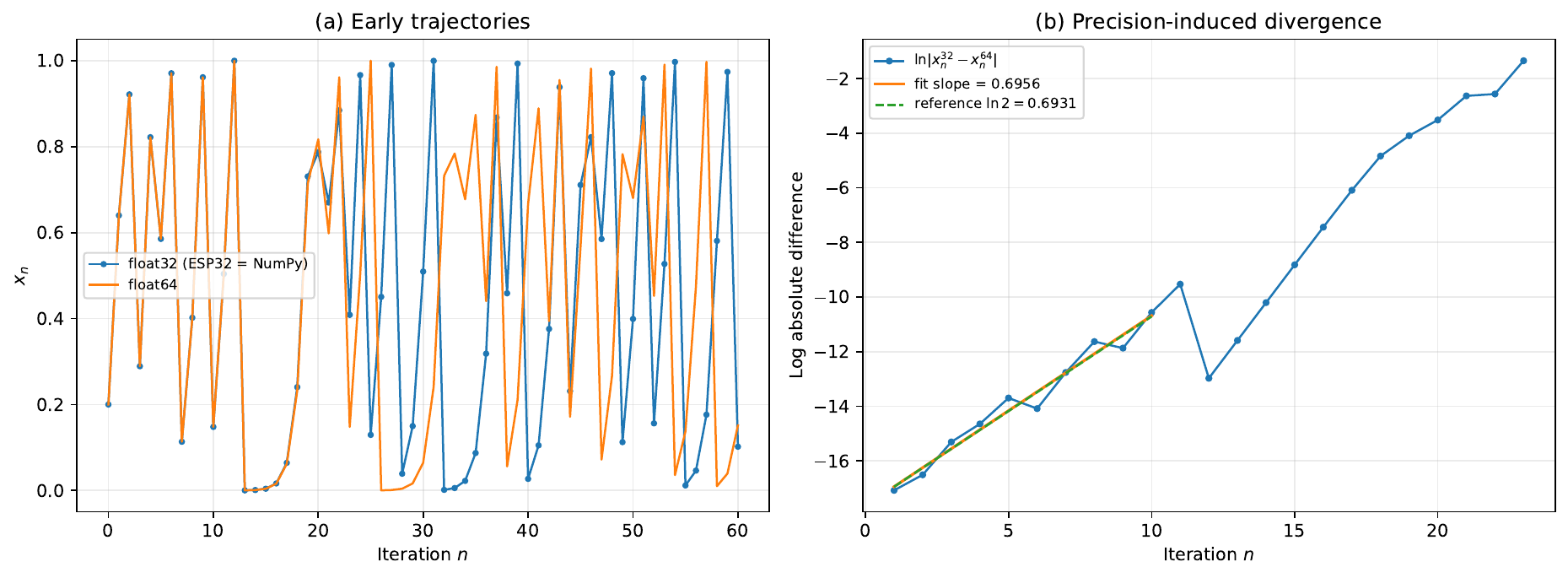}
\caption{Hardware validation and early precision sensitivity for the logistic map at $r=4$. (a) The ESP32 and NumPy \texttt{float32} realizations coincide bit-for-bit; the \texttt{float64} trajectory separates visibly after the first tens of iterations. (b) The early logarithmic \texttt{float32}--\texttt{float64} separation has slope $0.6956$, close to the real-valued LLE $\ln2=0.6931$.}
\label{fig:precision}
\end{figure}

\begin{revision}
By contrast, the \texttt{float64} trajectory begins to differ from \texttt{float32} immediately because the arithmetic precision is different. The early discrepancy is exponentially amplified. Fitting $\ln|x_n^{32}-x_n^{64}|$ over iterations $n=1,\ldots,10$ gives a slope of $0.695625$, only about $0.36\%$ above $\ln2$. This early growth also provides a direct numerical connection to the trajectory-fidelity scale in Eq.~\eqref{eq:Ts}. Converting the measured slope to base-10 units gives $\lambda_{10}=0.695625/\ln 10=0.302106$. For the logistic-map interval $D=1$ and the \texttt{float32} unit-roundoff scale $u=2^{-24}$, Eq.~\eqref{eq:Ts} predicts $T_s\approx22.9$ iterations; using the exact $\lambda=\ln2$ gives $T_s=23$. Thus the observed precision divergence occurs on the expected few-tens-of-iterations scale, long before exact digital recurrence. The close agreement of the early slope with $\ln2$ is consistent with previous work using numerical pseudo-orbit differences to access positive Lyapunov growth \cite{Peixoto2018,Zhou2019}; here it serves as a validation rather than as a new LLE estimator.

At much longer times the same \texttt{float32} control trajectory ceases to behave as an aperiodic chaotic record. The preperiodic transient lasts $\mu=1744$ iterations; after this transient the trajectory enters an exact digital cycle with period $P=4344$. Consequently, the first return to a previously visited machine state occurs at $\mu+P=6088$:
\begin{equation}
 \mu=1744, \qquad P=4344, \qquad \mu+P=6088.
\end{equation}
From that point onward, every subsequent value follows the same period-$4344$ digital cycle. Thus a million-point record contains an early region that displays the expected Lyapunov sensitivity but, after only a few thousand iterations, overwhelmingly consists of exact repetitions of a finite machine cycle (see Fig.~\ref{fig:precision}).
\end{revision}

\subsection{Exact periodicization is typical, not exceptional}

The control cycle is not a special feature of $x_0=0.2$. Every one of the 10,000 tested random \texttt{float32} initial conditions reached an exact repeated state. Only six digital periods were observed in this sweep (Table~\ref{tab:periods}). The period-$4344$ cycle dominated, attracting $67.24\%$ of the initial conditions. A period-1 state, corresponding to collapse to an exact fixed machine value, accounted for $19.19\%$. Periods 836 and 436 accounted for another $10.62\%$ and $2.83\%$, respectively; periods 136 and 143 were rare.

\begin{table}[t]
\centering
\caption{Exact digital-period distribution for 10,000 random \texttt{float32} initial conditions at $r=4$.}
\label{tab:periods}
\begin{tabular}{rrr}
\toprule
Period $P$ & Count & Fraction (\%) \\
\midrule
1    & 1919 & 19.19 \\
136  & 9    & 0.09 \\
143  & 3    & 0.03 \\
436  & 283  & 2.83 \\
836  & 1062 & 10.62 \\
4344 & 6724 & 67.24 \\
\bottomrule
\end{tabular}
\end{table}

The transient length $\mu$ had median 790 and maximum 3268. More directly relevant to long-record information content, the first exact repeat index $\mu+P$ had median 4651 and maximum 7612. Therefore, in this 10,000-seed experiment, no \texttt{float32} trajectory contributed a new state beyond 7612 iterations before entering exact repetition. A nominal record of one million points can consequently be dominated at the $>99\%$ level by repeated traversal of a finite digital cycle rather than by exploration of new states.

\begin{figure}[t]
\centering
\includegraphics[width=0.98\linewidth]{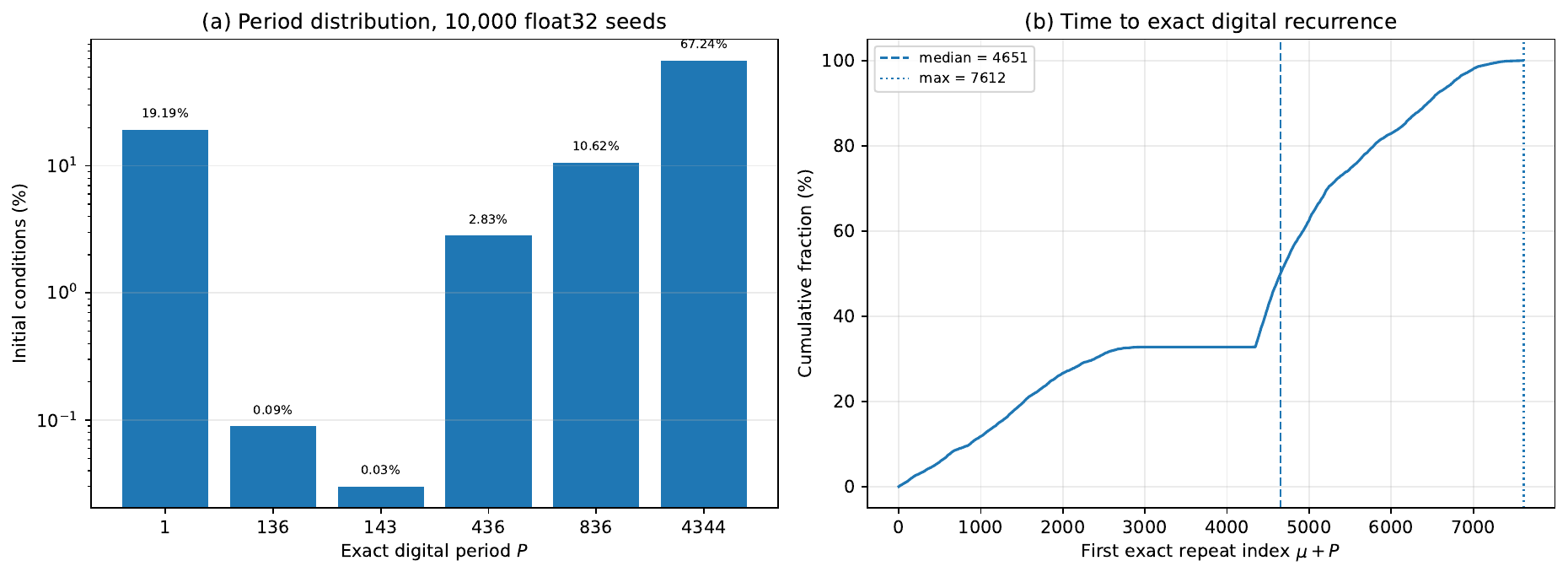}
\caption{Finite-precision periodicization across 10,000 random \texttt{float32} initial conditions. (a) The digital state space is dominated by a small number of exact cycles, especially $P=4344$. (b) The cumulative distribution of the first exact repeat index $\mu+P$ shows that all tested trajectories repeat by iteration 7612 (median 4651).}
\label{fig:cycles}
\end{figure}

The independent hardware set was consistent with this structure: all 130 valid long ESP32 records entered exact cycles, with the same dominant period families. Because the systematic software implementation had already been validated bit-for-bit against the ESP32 control trajectory, the larger 10,000-seed sweep is used for the statistics below (see Fig.~\ref{fig:cycles}).

\subsection{A long float32 record drives the forecast-error LLE to zero}

\begin{figure}[t]
\centering
\includegraphics[width=0.99\linewidth]{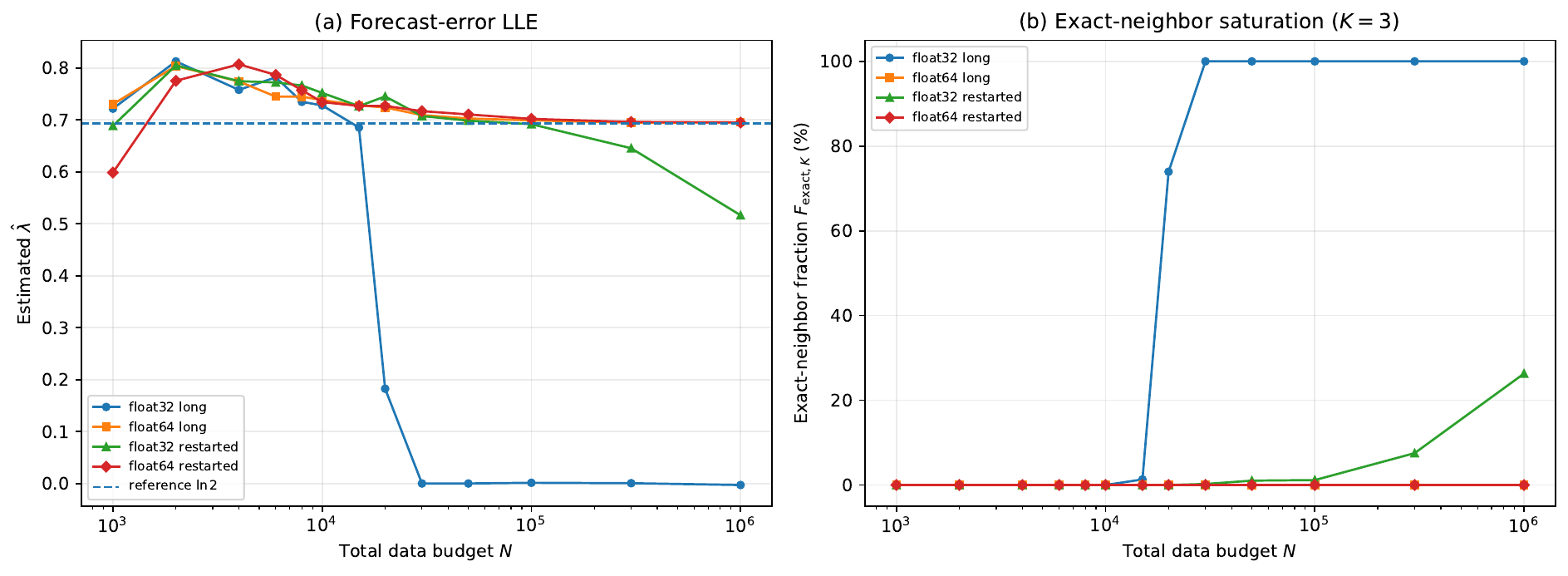}
\caption{Central experiment: data-budget dependence of the forecast-error LLE. (a) A single long \texttt{float32} record exhibits a sharp collapse from a positive LLE estimate to approximately zero, while the \texttt{float64} controls remain near $\ln2$. Restarted \texttt{float32} trajectories delay the degradation. (b) Exact-neighbor fraction $F_{\mathrm{exact},K}$, defined as the fraction of test histories for which all $K=3$ nearest neighbors in the training set are exact zero-distance matches. The rapid increase of $F_{\mathrm{exact},K}$ toward unity coincides with the long-record \texttt{float32} LLE collapse.}
\label{fig:collapse}
\end{figure}

Figure~\ref{fig:collapse}a shows the main result. At small and intermediate data budgets, the long-record \texttt{float32} estimate is close to the positive reference value, with ordinary finite-sample variability. For example,
\begin{equation}
\lamhat_{32,\mathrm{long}}(15{,}000)=0.6853,
\end{equation}
which is close to $\ln2$. Increasing the same record by only a few thousand points causes a sharp change:
\begin{align}
\lamhat_{32,\mathrm{long}}(20{,}000)&=0.1827,\\
\lamhat_{32,\mathrm{long}}(30{,}000)&=0.000245.
\end{align}
For $N\geq30{,}000$ the estimator remains essentially at zero, reaching $-0.0027$ at $N=10^6$.

This collapse is directly synchronized with the appearance of multiple exact neighbors in the training set (Fig.~\ref{fig:collapse}b). At $N=15{,}000$, only $1.31\%$ of queried test histories have all three nearest neighbors at exactly zero distance. At $N=20{,}000$, this fraction jumps to $73.94\%$, and at $N=30{,}000$ it reaches $100\%$. Once three exact copies of the same digital-cycle history are present in training, all three neighbors have the same deterministic future as the test history. The kNN average therefore reproduces the future exactly, $e_i(h)=0$, and the GMAE is pinned to the numerical floor across all forecast horizons. The line in Eq.~\eqref{eq:slope} becomes flat and $\lamhat\rightarrow0$.

The matched \texttt{float64} long-record control does not show this behavior on the tested scale. Its exact-neighbor fraction remains zero and the estimate converges toward the real-valued reference, reaching $0.6991$ at $N=10^5$ and $0.6953$ at $N=10^6$.

\subsection{\textcolor{black}{Information saturation and two mechanistic controls}}
\begin{figure}[H]
\centering
\includegraphics[width=0.49\linewidth]{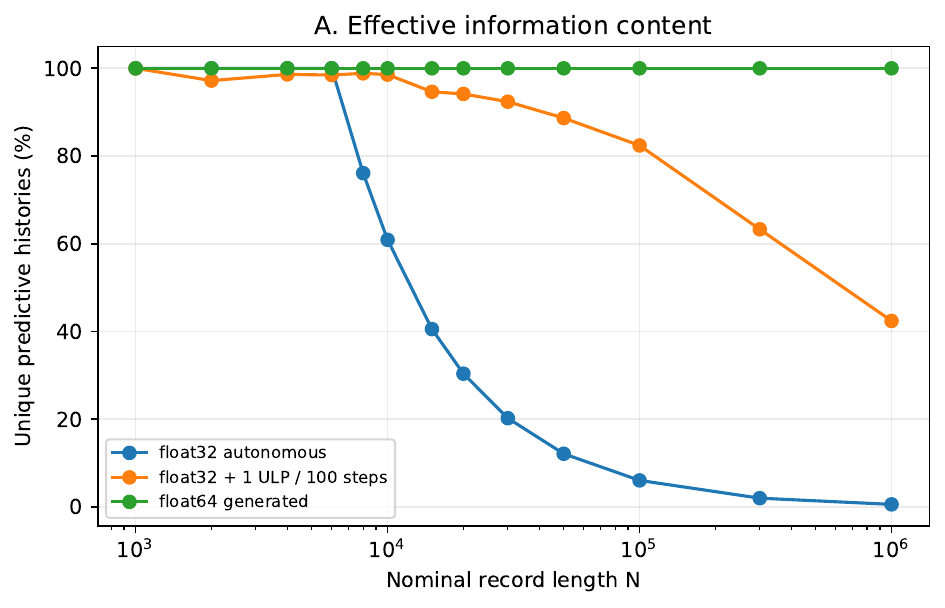}
\hfill
\includegraphics[width=0.49\linewidth]{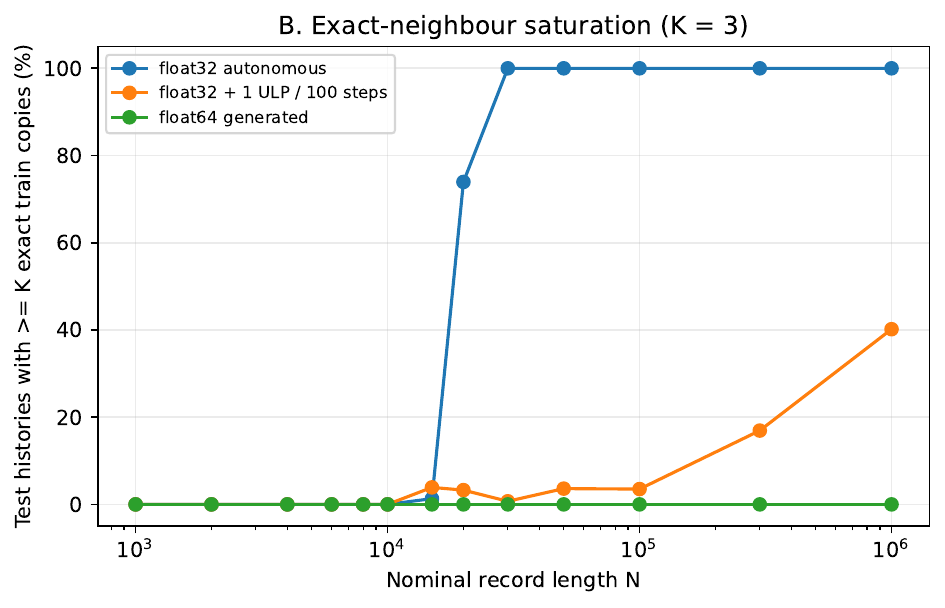}\\[0.5em]
\includegraphics[width=0.72\linewidth]{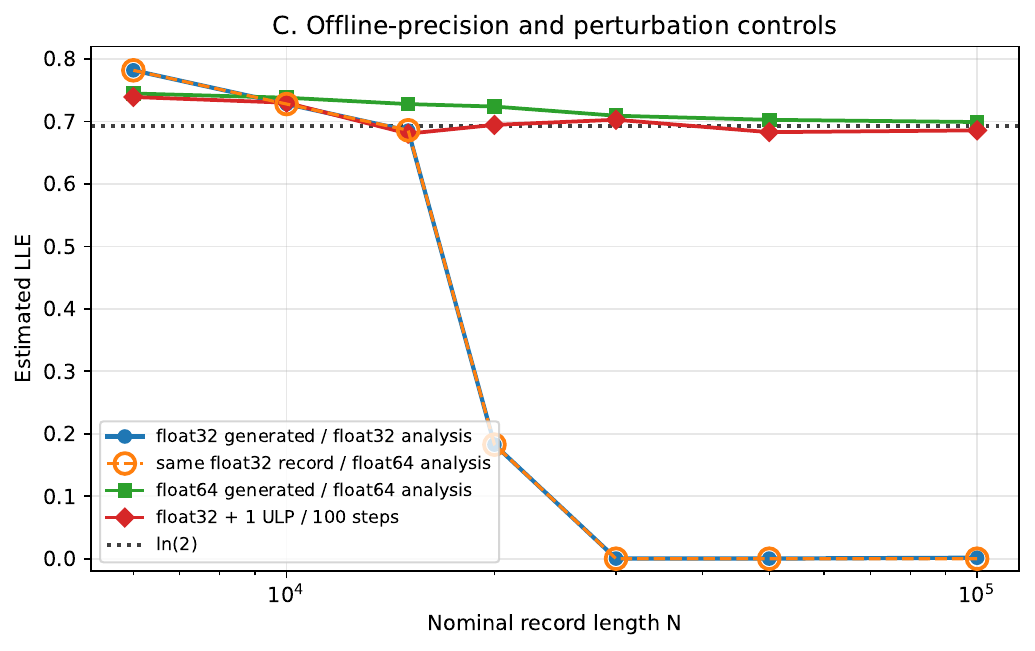}
\begin{revision}
\caption{Mechanistic controls linking nominal record length to effective predictive information. (a) Fraction of distinct predictive histories. The autonomous \texttt{float32} record saturates at a finite catalogue while $N$ continues to grow. (b) Fraction of test histories having at least $K=3$ exact training copies. This multiplicity, rather than the mere existence of one repeated history, tracks collapse of the kNN forecast-error LLE. (c) Offline-precision and perturbation controls. The \texttt{float32}-analysis curve and the post-cast \texttt{float64}-analysis curve for the same stored \texttt{float32} record nearly overlap, demonstrating that higher-precision offline analysis does not restore information lost during generation. A separately generated \texttt{float64} trajectory remains near $\ln2$, while a one-ULP perturbation every 100 generation steps prevents fixed-cycle domination and preserves a positive estimate over the tested range.}
\label{fig:mechanisticcontrols}
\end{revision}
\end{figure}

\begin{revision}
Using the unique-history fraction $R_H(N)$ defined in Eq.~\eqref{eq:uniquehist} makes the phrase ``more data become less informative'' quantitative. For the autonomous \texttt{float32} control trajectory, $R_H=100\%$ at $N=6000$, so all admissible predictive histories are distinct. After the first exact recurrence, however, the number of distinct histories rapidly saturates at approximately 6068 even while the nominal record continues to grow. Consequently, the unique-history fraction $R_H$ decreases from 60.91\% at $N=10{,}000$ to 30.40\% at $N=20{,}000$, 6.07\% at $N=100{,}000$, and only 0.61\% at $N=10^6$ [Fig.~\ref{fig:mechanisticcontrols}(a)]. The separately generated \texttt{float64} trajectory remains essentially unique over this range, whereas the periodically perturbed control loses uniqueness much more slowly.

The stronger multiplicity criterion explains why the existence of a single repeated history is not sufficient by itself to produce collapse. In the autonomous \texttt{float32} record, the fraction of test histories with at least $K=3$ exact training copies increases from 1.31\% at $N=15{,}000$ to 73.94\% at $N=20{,}000$ and 100\% at $N=30{,}000$. By contrast, the one-ULP perturbation control has only 3.27\%, 0.71\%, and 3.51\% such histories at $N=20{,}000$, $30{,}000$, and $100{,}000$, respectively [Fig.~\ref{fig:mechanisticcontrols}(b)], while its LLE estimate remains positive. Thus, the relevant mechanism is not merely the presence of occasional duplicates but saturation by enough exact predictive copies to dominate the $K$-neighbor forecast.

The offline-precision control directly tests whether moving the analysis from a microcontroller record to higher-precision desktop arithmetic can recover the lost instability information. It cannot. At $N=20{,}000$, the same \texttt{float32}-generated trajectory gives $\lamhat=0.182654$ when analyzed as \texttt{float32} and $0.182616$ after casting to \texttt{float64}. At $N=30{,}000$, the two estimates are $0.000245$ and numerically zero, respectively. At $N=100{,}000$, the cast-to-\texttt{float64} analysis remains at numerical zero, whereas a trajectory generated from the outset in \texttt{float64} gives $\lamhat=0.699104$ [Fig.~\ref{fig:mechanisticcontrols}(c)]. Higher precision during post-processing therefore does not reconstruct states or futures that were already duplicated during reduced-precision generation.

The scheduled one-ULP control gives the complementary result. Shifting the state by the smallest representable \texttt{float32} increment every 100 iterations prevents one autonomous digital cycle from dominating the forecast catalogue. The estimated LLE remains $0.694510$ at $N=20{,}000$, $0.702756$ at $N=30{,}000$, and $0.685754$ at $N=100{,}000$, all close to $\ln2$. This verifies that occasional perturbations are an effective established mitigation of fixed-cycle trapping, while also showing that they answer a different question from the restarted protocol: the perturbation control modifies the generated dynamics, whereas restarting changes how unperturbed trajectory segments are acquired.
\end{revision}

\subsection{Restarted short trajectories preserve dynamical information longer}

\rev{The two acquisition protocols defined in Section~\ref{sec:protocols} can now be compared at the same total data budget. At $N=100{,}000$, the long-record protocol contains one continuous \texttt{float32} trajectory, whereas the restarted protocol contains 1000 independently initialized trajectories of length 100.} The continuous record is already deep in its repeated digital cycle and gives
\begin{equation}
 \lamhat_{32,\mathrm{long}}=0.0016.
\end{equation}
Distributing the same 100,000 stored points among 1000 independently restarted trajectories of length 100 gives
\begin{equation}
 \lamhat_{32,\mathrm{restart}}=0.6917,
\end{equation}
which is very close to $\ln2$. The corresponding \texttt{float64} values are $0.6991$ and $0.7019$, respectively (Table~\ref{tab:budget100k}). Thus, at a matched data budget, the loss of the LLE is not caused by an insufficient number of samples; it is caused by how the finite-precision samples revisit the same deterministic states.

\begin{table}[t]
\centering
\caption{Matched $N=100{,}000$ data budget. The underlying estimator and split rule are unchanged; only precision and data organization differ.}
\label{tab:budget100k}
\begin{tabular}{lcc}
\toprule
Condition & $\lamhat$ & $\fexact$ (\%) \\
\midrule
\texttt{float32}, long      & 0.0016 & 100.00 \\
\texttt{float32}, restarted & 0.6917 & 1.17 \\
\texttt{float64}, long      & 0.6991 & 0.00 \\
\texttt{float64}, restarted & 0.7019 & 0.00 \\
Reference real-valued map   & 0.6931 & -- \\
\bottomrule
\end{tabular}
\end{table}

Restarting is not an unlimited cure. At very large total budgets, independently restarted \texttt{float32} trajectories themselves begin to collide in the finite state space. For the restarted protocol, $\fexact$ rises from $1.17\%$ at $N=10^5$ to $7.52\%$ at $N=3\times10^5$ and $26.31\%$ at $N=10^6$. In parallel, the estimated LLE decreases from $0.6917$ to $0.6454$ and $0.5164$. Restarting therefore \emph{delays} finite-state saturation by preventing one short cycle from dominating the entire record, but it cannot make the finite state space infinite.

\subsection{Collapse scale follows the digital period across initial conditions}

The multi-seed experiment confirms that the phenomenon is not tied to the control initial condition. All 28 representative \texttt{float32} seeds, spanning six exact digital periods and a broad range of transient lengths, reached a regime with $\fexact\geq0.5$ within the tested budgets. All 28 also reached $|\lamhat|\leq0.1$. At the maximum tested budget $N=100{,}000$, the mean \texttt{float32} estimate across these deliberately stratified seeds is $0.00024$ (range $-0.00468$ to $0.00478$), whereas the matched \texttt{float64} mean is $0.69749$ (range $0.68430$ to $0.70575$), with zero exact-$K$ saturation in all \texttt{float64} cases.

\begin{figure}[t]
\centering
\includegraphics[width=0.99\linewidth]{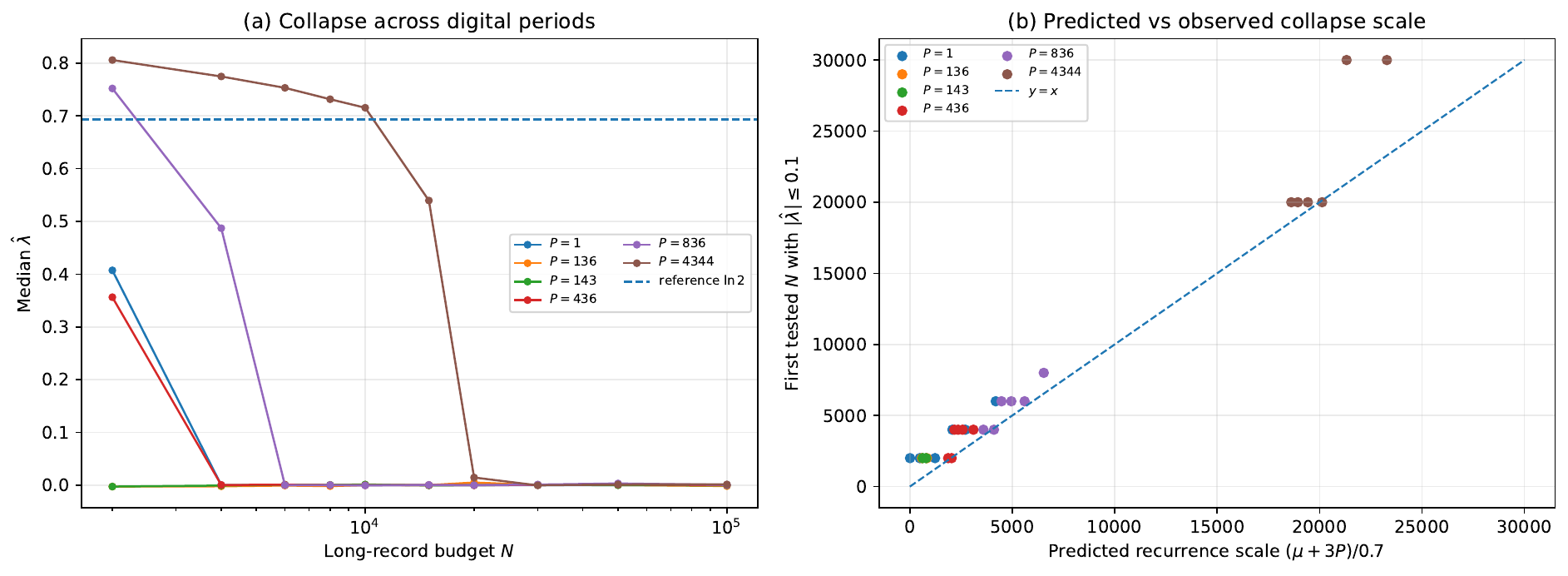}
\caption{Generality across initial conditions. (a) Median \texttt{float32} LLE estimates grouped by exact digital period. Shorter cycles lead to earlier collapse. The 28-seed set is stratified by period and therefore is not intended to represent natural basin probabilities. (b) The observed collapse scale tracks the interpretable recurrence scale $(\mu+3P)/0.7$.}
\label{fig:generality}
\end{figure}

Importantly, the position of the collapse changes systematically with the digital period. Short cycles such as $P=436$ and $P=836$ cause much earlier loss of the positive forecast-error slope than the dominant $P=4344$ cycle (Fig.~\ref{fig:generality}a). The simple recurrence scale from Eq.~\eqref{eq:predscale},
\begin{equation}
 N_{3}^{\mathrm{pred}}\approx\frac{\mu+3P}{0.7},
\end{equation}
strongly correlates with the first tested data budget at which $|\lamhat|\leq0.1$: Pearson $r=0.982$ and Spearman $\rho=0.969$. The same predicted scale correlates with the onset of $\fexact\geq0.5$ with Pearson $r=0.985$ and Spearman $\rho=0.959$. These correlations are not introduced as a universal law, because the exact transition also depends on history length, neighbor multiplicity, split geometry, and sampling. They nevertheless demonstrate that the estimator collapse is controlled by the finite-state recurrence structure rather than by a peculiar property of one seed.

\begin{revision}
Table~\ref{tab:periodfamilies} summarizes one representative seed near the central part of the selected transient-length span for each of the six observed period families. The columns report the measured transient $\mu$, first exact recurrence $\mu+P$, the recurrence-occupancy scale $N_3^{\mathrm{pred}}=(\mu+3P)/0.7$, and the first tested data budget for which $|\widehat{\lambda}|\leq0.1$. For the three shortest-period families, the predicted transition lies below the minimum tested budget of 2000 points and the estimate is already collapsed at that first budget. For $P=436$, $836$, and $4344$, the transition moves progressively to larger records, directly showing that different reached periodic orbits produce different collapse onsets.

\begin{table}[t]
\centering
\color{black}
\small
\caption{Representative logistic-map period families from the 28-seed stratified analysis. $N_{\rm coll}$ is the first tested budget satisfying $|\widehat{\lambda}|\leq0.1$; ``$\leq2000$'' indicates that the estimate was already collapsed at the smallest tested budget. The table is illustrative of the period dependence and is not a basin-probability sample.}
\label{tab:periodfamilies}
\begin{tabular}{rrrrrr}
\toprule
$P$ & $\mu$ & $\mu+P$ & $N_3^{\rm pred}$ & $N_{\rm coll}$ & $\widehat{\lambda}(N_{\rm coll})$ \\
\midrule
1    & 857  & 858  & 1,229  & $\leq2,000$ & $\approx0$ \\
136  & 171  & 307  & 828    & $\leq2,000$ & $-0.0056$ \\
143  & 122  & 265  & 788    & $\leq2,000$ & $-0.0056$ \\
436  & 213  & 649  & 2,173  & 4,000         & $0.00075$ \\
836  & 622  & 1,458& 4,472  & 6,000         & $-0.00215$ \\
4,344& 1,055& 5,399& 20,125 & 20,000        & $0.0321$ \\
\bottomrule
\end{tabular}
\end{table}
\end{revision}

\subsection{\textcolor{black}{Predictor dependence: recursive local-linear control}}
\begin{figure}[t]
\centering
\includegraphics[width=0.99\linewidth]{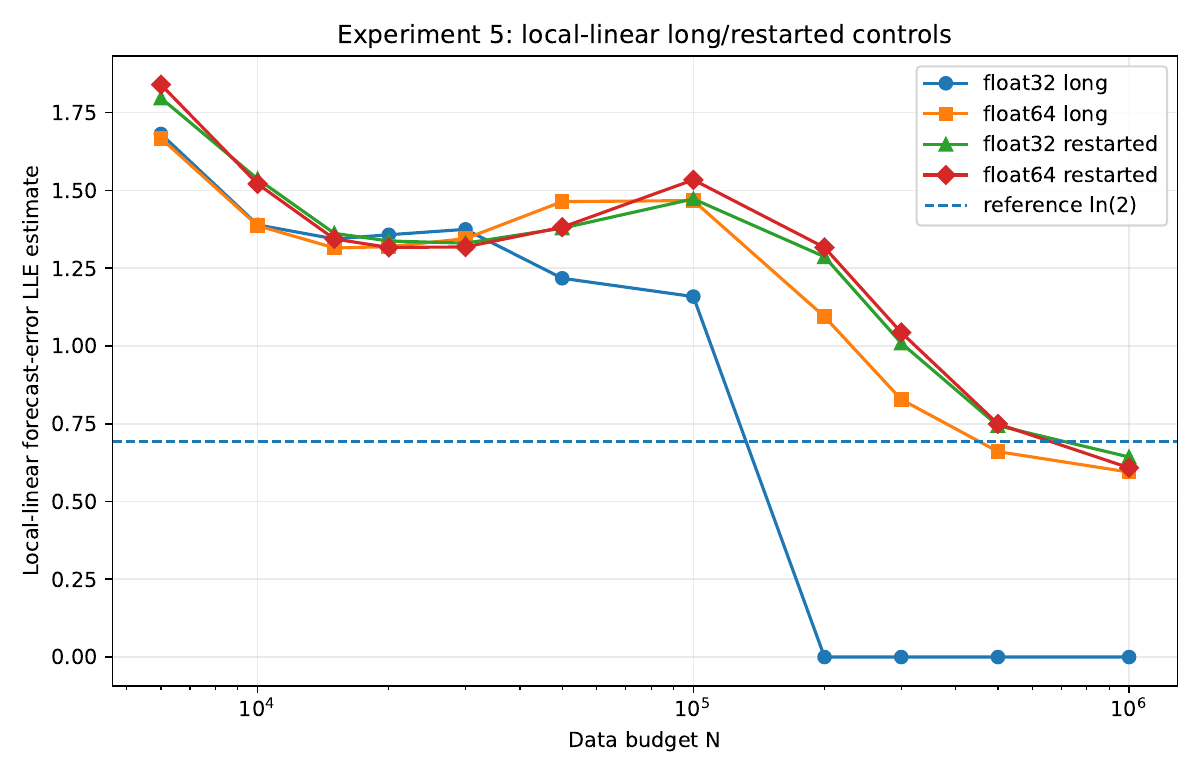}
\begin{revision}
\caption{Recursive local-linear predictor control. A one-step local-affine model with $K_{\mathrm{LL}}=20$ neighbors is recursively iterated to forecast horizons 1--5. The autonomous long \texttt{float32} record shows an abrupt suppression of forecast-error growth to numerical zero at $N=200{,}000$, where all 20 nearest training neighbors are exact copies. Long \texttt{float64} and independently restarted \texttt{float32}/\texttt{float64} controls remain positive. Absolute local-linear slopes at smaller $N$ are model dependent and are used here as a robustness control rather than as an independently calibrated LLE estimator.}
\label{fig:locallinearcontrol}
\end{revision}
\end{figure}

\begin{revision}
Replacing local-constant kNN averaging by a recursively iterated local-affine predictor does not remove the finite-state failure mode, but it changes the data budget at which the failure becomes complete. With $K_{\mathrm{LL}}=20$, the long \texttt{float32} forecast-error slope remains positive through $N=100{,}000$ ($\widehat{\lambda}_{\mathrm{LL}}=1.1587$), even though the original $K=3$ kNN estimator has already collapsed by $N=30{,}000$. At $N=100{,}000$, the local-linear neighborhood contains on average 15.71 exact copies among its 20 nearest neighbors, but no queried history yet has all 20 neighbors at zero distance, so $F_{\mathrm{exact},20}=0$.

The transition becomes complete at $N=200{,}000$. At this budget every queried long-\texttt{float32} history has 20 exact training copies ($F_{\mathrm{exact},20}=1$), the recursive forecasts become exact to the numerical floor over all five horizons, and the fitted growth slope is numerically zero. The long-\texttt{float32} local-linear slope remains at numerical zero for all larger tested budgets up to $10^6$. In contrast, at $N=200{,}000$ the matched long-\texttt{float64}, restarted-\texttt{float32}, and restarted-\texttt{float64} controls remain positive at 1.0939, 1.2867, and 1.3165, respectively [Fig.~\ref{fig:locallinearcontrol}]. Even at $N=10^6$, the corresponding control slopes remain positive (0.5945, 0.6430, and 0.6082), whereas the autonomous long-\texttt{float32} value is still zero.

The absolute local-linear slopes are systematically above $\ln2$ at smaller data budgets, reflecting model- and sample-dependent approximation error; therefore this arm is not presented as a second calibrated LLE estimator. Its purpose is mechanistic. The result shows that forecast-error suppression is not unique to a $K=3$ local-constant predictor: a different local forecasting architecture also loses growth once its own larger neighborhood becomes fully saturated by exact periodic copies. At the same time, the later transition demonstrates that the precise collapse threshold is predictor dependent.
\end{revision}

\subsection{\textcolor{black}{Generality beyond the logistic $r=4$ case: H\'enon map}}
\begin{figure}[H]
\centering
\includegraphics[width=0.99\linewidth]{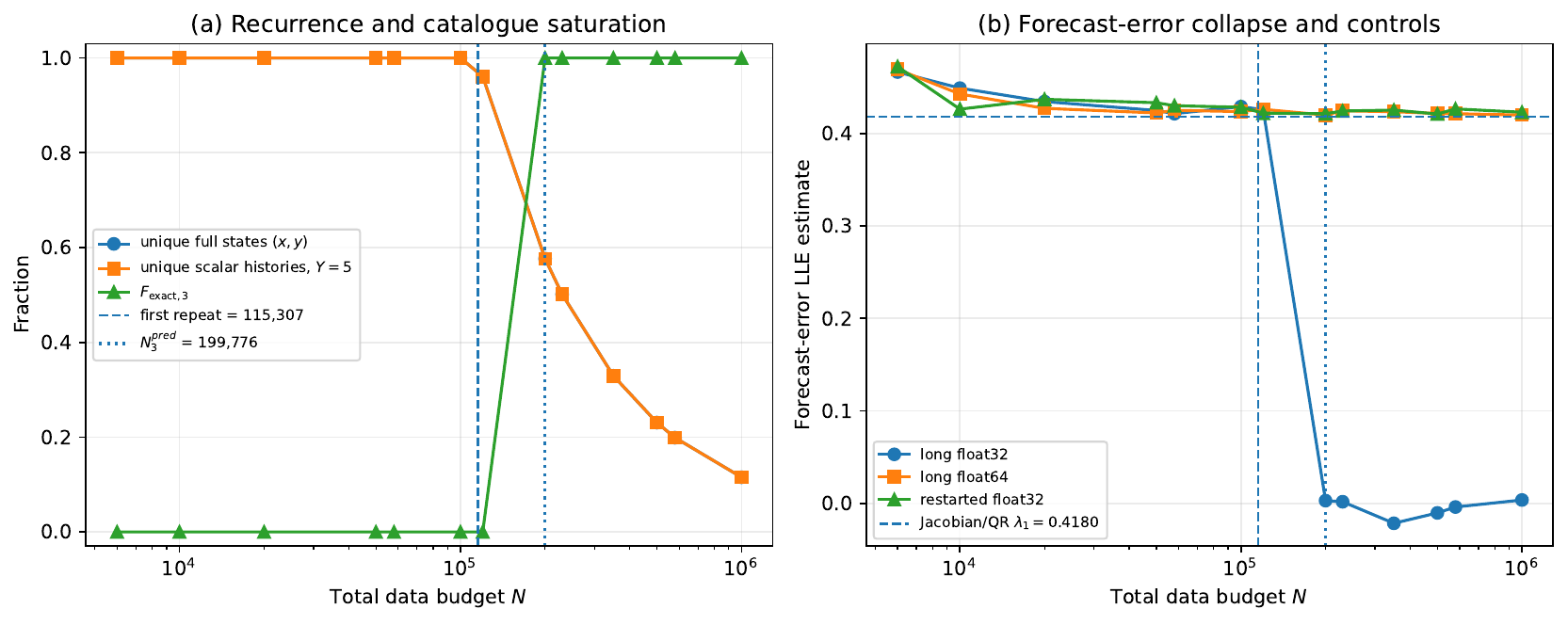}
\begin{revision}
\caption{H\'enon-map generalization for $a=1.4$, $b=0.3$. (a) Exact recurrence of the full \texttt{float32} state pair begins at $\mu+P=115{,}307$, after which the numbers of distinct states and scalar $Y=5$ histories saturate. The vertical dotted line marks the parameter-free $K=3$ catalogue scale $N_3^{\mathrm{pred}}=(\mu+3P)/0.7\approx199{,}776$, where $F_{\mathrm{exact},3}$ reaches unity. (b) The long-\texttt{float32} forecast-error LLE remains close to the independent Jacobian/QR reference $\lambda_1=0.41801$ after the first recurrence, but collapses near zero when the $K=3$ catalogue becomes fully saturated. Long \texttt{float64} and restarted \texttt{float32} controls remain close to the reference throughout the tested range.}
\label{fig:henon}
\end{revision}
\end{figure}

\begin{revision}
The H\'enon control reproduces the same recurrence-to-saturation mechanism in a two-dimensional map without the special binary structure of the logistic $r=4$ update. The independent tangent-space calculation gives
\begin{equation}
 \lambda_1=0.4180097921,\qquad \lambda_2=-1.6219825964,
\end{equation}
with $\lambda_1+\lambda_2=-1.2039728043=\ln(0.3)$ to within $1.5\times10^{-14}$. The largest exponent is also consistent with the literature value $0.418$ and the interval-extension estimate $0.417$ reported for the canonical H\'enon regime by Mendes and Nepomuceno \cite{Mendes2016}.

The autonomous \texttt{float32} H\'enon trajectory has a much longer digital transient than the logistic control but still enters an exact machine cycle. Bitwise recurrence of the full pair $(x,y)$ occurs with
\begin{equation}
 \mu=103{,}039,\qquad P=12{,}268,\qquad \mu+P=115{,}307.
\end{equation}
This first repeat does not by itself collapse the forecast estimator. At $N=120{,}000$, the number of distinct full states has already saturated at 115,307, yet $F_{\mathrm{exact},3}=0$ and the long-\texttt{float32} estimate remains $\lamhat=0.42553$, close to the independent reference [Fig.~\ref{fig:henon}(a,b)]. Thus exact recurrence and estimator failure remain distinct events.

The catalogue-multiplicity scale from Eq.~\eqref{eq:predscale} predicts
\begin{equation}
 N_{3}^{\mathrm{pred}}\approx
 \frac{103{,}039+3\times12{,}268}{0.7}
 =199{,}776.
\end{equation}
No parameter is fitted in this calculation. At the next tested budget, $N=200{,}000$, every queried long-\texttt{float32} history has at least three exact training copies ($F_{\mathrm{exact},3}=1$) and the forecast-error slope collapses to $0.00293$. The matched long-\texttt{float64} and restarted-\texttt{float32} estimates remain $0.41973$ and $0.42126$, respectively. At $N=10^6$ the autonomous long-\texttt{float32} value is still near zero ($0.00364$), whereas long \texttt{float64} and restarted \texttt{float32} remain at $0.41984$ and $0.42285$. The restarted arm has occasional single exact neighbors at large $N$ but never reaches three exact copies for the queried histories over the tested range.

These results provide a cross-map validation of the central mechanism. The numerical recurrence scales differ by more than an order of magnitude between the logistic and H\'enon maps, but in both systems the positive forecast-error slope is retained after the first exact repeat and is lost only when recurrence has populated the training catalogue with the multiplicity of exact futures required by the predictor.
\end{revision}

\section{Discussion}
\label{sec:discussion}

\begin{revision}
\subsection{Mechanism and separation of finite-precision scales}

The experiments separate three finite-precision phenomena that are often discussed together but occur on very different scales. At short times, a microscopic numerical discrepancy grows at approximately the positive Lyapunov rate. Equation~\eqref{eq:Ts} places this pointwise trajectory-fidelity scale at about 23 iterations for \texttt{float32} and 52 iterations for \texttt{float64} in the logistic $r=4$ benchmark when $D=1$ and $\lambda=\ln2$. These values concern loss of pointwise numerical fidelity, not exact state recurrence. In the ESP32-validated \texttt{float32} control trajectory, the first exact repeated state occurs much later at $\mu+P=6088$. Even that recurrence is not yet the estimator-failure threshold: the $K=3$ forecast-error estimate remains positive until repeated traversals of the digital cycle supply enough exact training copies to saturate the predictor's neighborhood. The H\'enon experiment shows the same ordering on a very different numerical scale: full-state recurrence begins at 115,307 samples, while the forecast-error estimate is still $0.4255$ at $N=120{,}000$ and collapses only near $N=200{,}000$.

This separation clarifies the relation to prior work. Finite-precision periodicization, state-graph degradation, and methods for disrupting short digital cycles are established topics \cite{Persohn2012,Galias2021,Antonelli2018,Kloewer2023,Fan2021,Merah2021,Li2024}. Lower-bound-error studies likewise address numerical pseudo-orbit reliability and Lyapunov growth \cite{Mendes2016,Nepomuceno2017,Nepomuceno2018}. The contribution here is not the existence of digital cycles, but the later transition by which recurrence changes the statistical learning problem itself: the forecast catalogue becomes populated by duplicate histories carrying duplicate deterministic futures. The estimator can therefore remain useful after the first exact repeat and fail only when the predictor's required local multiplicity is saturated.

The term ``collapse'' is correspondingly narrow. It denotes suppression of the \emph{forecast-error growth estimate} by catalogue saturation; it does not mean that the underlying mathematical LLE becomes zero or that unrelated LLE estimators must fail. The H\'enon Jacobian/QR reference remains at $0.41801$ while the forecast-error estimate collapses, providing an internal counterexample to any broader interpretation. LBE/interval-extension methods use pseudo-orbit separation rather than a predictive catalogue, and recent work has reported successful LBE-based experimental LLE estimates for physical Jerk and Chua circuits \cite{Nazare2026}. Classical tangent-space and other non-catalogue estimators can have their own numerical limitations, but the duplicate-history mechanism identified here should not be projected onto them without separate tests.

\subsection{Equation-free relevance and effective information content}

This methodological specificity does not reduce the practical relevance of the problem because model-based and equation-free approaches address complementary settings. Jacobian/QR analysis presupposes access to governing equations or a differentiable dynamical model, and the LBE/interval-extension construction likewise exploits known equations of motion \cite{Mendes2016}. In experimental, biomedical, environmental, and archived-signal applications, however, the analyst may have only a scalar measured time series, with no trusted equations, tangent model, or full state vector. Forecast-error profiling is designed for precisely this equation-free setting \cite{Velichko2026FEGPro,Velichko2026Negative}. In such applications, an independent equation-based Lyapunov calculation may simply not be available as a replacement or cross-check.

The unique-history fraction $R_H(N)$ makes the information loss explicit. Nominal sample count continues to increase while the number of distinct predictive histories approaches a finite ceiling; in the autonomous logistic \texttt{float32} record, only about 0.61\% of admissible histories remain unique at $N=10^6$. This is not ordinary statistical overfitting. Before saturation, more samples improve local coverage. After saturation, coverage becomes perfect for the wrong reason: new samples increasingly reproduce history--future pairs already stored in the catalogue. Sample count and effective dynamical diversity are therefore not interchangeable measures of information content.

The implication is broader than one scalar LLE-like slope. FEG-Pro retains the multi-horizon forecast-error curve as a structured profile from which local slopes, curvature, residual roughness, monotonicity, fit quality, and forecast-error distribution entropy can be extracted \cite{Velichko2026FEGPro}. Forecast-error-growth quantities have already been evaluated as exploratory nonlinear predictability features in real respiratory recordings \cite{VelichkoHuyut2027}; related work uses short-trajectory forecast-error contraction for equation-free negative-LLE estimation \cite{Velichko2026Negative}, and a separate screening framework compares kNN multi-horizon profiles with Mittag--Leffler-compatible and conventional alternatives \cite{NGbo2026Mittag}. The saturation mechanism acts upstream of these summaries because it changes the empirical error profile itself. The present study directly establishes suppression of the growth slope and distortion of the underlying profile; it does \emph{not} assume that every downstream descriptor is affected identically. Quantifying feature-specific sensitivity is a separate problem.

The same quantities can also serve as diagnostics of information saturation. A systematic decrease of forecast-error growth accompanied by increasing exact-neighbor multiplicity, such as $F_{\mathrm{exact},K}$, indicates that newly acquired data are increasingly reproducing previously observed predictive patterns. This signature is not unique to strict digital periodicity--stable dynamics, leakage, noise floors, or other forms of regularity can also flatten forecast errors--so it should be interpreted as a recurrence/information-saturation diagnostic rather than as a standalone periodicity test.

\subsection{Generation precision, perturbation, and acquisition strategy}

The ESP32 comparison shows that generation precision is part of the effective dynamical model rather than merely an implementation detail. NumPy \texttt{float32} reproduces the tested hardware trajectory bit-for-bit, so the later software sweeps characterize the same finite-state realization. Once a reduced-precision generator has produced exact duplicate histories, transferring the stored record to a desktop and casting it to \texttt{float64} cannot restore missing states or split identical histories. The near-complete overlap of the \texttt{float32}-analysis and post-cast \texttt{float64}-analysis curves in Fig.~\ref{fig:mechanisticcontrols}(c) demonstrates this directly. By contrast, a trajectory generated in \texttt{float64} remains close to the positive reference over the tested budgets.

The one-ULP perturbation provides a causal test of the proposed mechanism. Perturbation and parameter-variation strategies for avoiding short-period digital regimes are already established \cite{Tutueva2020}; the intervention itself is not claimed as novel. Here, shifting the state by only one representable \texttt{float32} ULP every 100 iterations prevents domination by a single autonomous digital cycle and simultaneously prevents forecast-error collapse over the tested range. Low precision alone is therefore insufficient to produce the observed failure. The failure emerges when autonomous finite-precision recurrence is allowed to saturate the predictive catalogue with exact repeated histories and futures. The perturbation control changes the generator and makes it externally driven, which is why it answers a different question from restarted acquisition.

Restarting changes the data-acquisition protocol without perturbing the dynamics within any individual segment. At a matched $N=100{,}000$ budget, distributing the data among independently initialized length-100 trajectories changes the logistic \texttt{float32} estimate from essentially zero to $0.6917$, close to $\ln2$. The strategy works because each short trajectory repeatedly samples a pre-periodic transient rather than allowing one eventual cycle to dominate the entire data set. It is therefore particularly natural in simulations, repeatable experiments, and repeated-relaxation measurements where independent restarts are available. Short-trajectory ensembles also arise independently in equation-free contraction analysis \cite{Velichko2026Negative}. Restarting is not unlimited, however: at very large ensemble sizes, collisions occur across trajectories and exact neighborhoods reappear. The relevant concept is finite-state occupancy, not simply ``short is better than long.''

\subsection{Predictor dependence, map generality, and limitations}

The recursive local-linear control separates the data-level mechanism from the response of a particular predictor. Exact recurrence and saturation of distinct history--future pairs belong to the recorded trajectory and do not depend on kNN. The collapse threshold does depend on the forecast architecture. The original local-constant predictor requires only three exact neighbors and is suppressed near $N=30{,}000$, whereas the $K_{\mathrm{LL}}=20$ recursive local-affine predictor remains positive much longer and collapses only at $N=200{,}000$, when all 20 nearest neighbors are exact copies for every query. Thus the architecture changes the onset but not the route to failure once its effective local neighborhood becomes fully populated by repeated deterministic futures. The local-linear slopes are not independently calibrated as unbiased LLE estimates, and the result should not be extrapolated quantitatively to reservoir or neural forecasters without direct tests.

The H\'enon map provides the complementary dynamical-system control. It differs from logistic $r=4$ in dimension, algebraic structure, Lyapunov exponent, digital transient, and cycle period, yet the same ordering is observed: exact recurrence occurs first, and estimator collapse occurs later when the local predictive catalogue reaches the required multiplicity. The independently measured $\mu$ and $P$ give $N_3^{\mathrm{pred}}\approx199{,}776$, essentially the tested $N=200{,}000$ point where $F_{\mathrm{exact},3}=1$ and forecast-error growth disappears. Together with the 28-seed logistic analysis, this supports $(\mu+KP)/f_{\mathrm{train}}$ as an interpretable occupancy scale for the chronological local-neighbor construction, not as a universal law.

The scope remains deliberately limited. Two canonical maps and two local forecast architectures are sufficient to rule out an explanation tied only to the binary peculiarity of logistic $r=4$, but they do not establish universality across all finite-precision systems or all forecast models. Continuous-time flows, noisy experimental signals, fixed-point and half-precision arithmetic, and nonlocal forecasters are important next tests. Finally, the intended scientific object must be stated explicitly: after cycle capture, the asymptotic \texttt{float32} machine system is genuinely periodic, so a near-zero forecast-error slope is not incorrect if that finite-state automaton itself is the object of interest. It becomes misleading when it is interpreted as the instability of the intended real-valued chaotic system.
\end{revision}
\section{Conclusions}
\label{sec:conclusions}

\begin{revision}
This study identifies a specific reliability limit of equation-free forecast-error Lyapunov inference from long finite-precision records. Digital periodicity itself is well known; the key result is that the forecast-error estimate does not fail at the first exact recurrence. Failure occurs later, when repeated traversals populate the predictive catalogue with enough exact copies of the same histories and deterministic futures to suppress multi-horizon forecast-error growth. The separation between pointwise trajectory-fidelity loss, exact machine-state recurrence, and catalogue saturation is observed across logistic-map period families and is independently reproduced by the H\'enon map.

The controls localize the mechanism. Reanalyzing an already generated \texttt{float32} record in \texttt{float64} does not recover the lost information, whereas generation in \texttt{float64} remains non-collapsed over the tested range. A one-ULP intervention every 100 steps prevents fixed-cycle domination, providing a causal control that the failure is recurrence-mediated rather than an unavoidable consequence of single-precision arithmetic. Independently restarted short trajectories preserve useful dynamical information much longer without perturbing the autonomous dynamics within each segment. A recursive local-linear predictor reaches the same qualitative endpoint at a later threshold, showing that information saturation belongs to the data while the precise failure budget is predictor dependent.

The practical implication is that nominal record length is not a sufficient measure of information content in reduced-precision chaotic data. When governing equations and tangent dynamics are unavailable, forecast-error profiles can provide useful instability estimates and nonlinear predictive features, but long records should be checked for exact recurrence, predictive-history diversity, and exact-neighbor multiplicity before additional samples are assumed to add information. The present claim is restricted to forecast-error/analogue inference and does not imply a general failure of Lyapunov-exponent estimation. For equation-free applications, generation precision and finite-state occupancy should therefore be treated as part of the inference design, not merely as implementation details.
\end{revision}
\section*{CRediT authorship contribution statement}
\textbf{Andrei Velichko:} Conceptualization, Formal analysis, Funding acquisition, Methodology, Software, Supervision, Writing -- original draft, Writing -- review \& editing.
\textbf{Viet-Thanh Pham:} Data curation, Formal analysis, Investigation, Validation, Writing -- original draft, Writing -- review \& editing.

\section*{Declaration of competing interest}
The authors declare that they have no known competing financial interests or personal relationships that could have appeared to influence the work reported in this paper.

\section*{Supplementary Material}
\rev{The Supplementary Material is organized as a reproducibility package for the principal numerical results. It includes the baseline logistic-map and ESP32 validation code, information-saturation and precision/perturbation controls, representative period-family data, the recursive local-linear predictor control, and the H\'enon recurrence/QR/forecast-error analysis. Each component is accompanied by concise reproduction instructions and machine-readable numerical outputs used for the corresponding figures and tables.}

\section*{Data and code availability}
The Python code and machine-readable outputs underlying the main numerical experiments are provided with this article as Supplementary Material. The ESP32 validation material and the scripts required to reproduce the information-saturation, offline-precision, one-ULP perturbation, local-linear, and H\'enon controls are included in the same package. Additional supporting hardware data are available from the corresponding author upon reasonable request.

\section*{Funding}
This research was funded by the Russian Science Foundation, grant number 22-11-00055-P.


\begin{thebibliography}{99}

\bibitem{Wolf1985}
A. Wolf, J.B. Swift, H.L. Swinney, J.A. Vastano,
Determining Lyapunov exponents from a time series,
\textit{Physica D: Nonlinear Phenomena} 16 (3) (1985) 285--317.
https://doi.org/10.1016/0167-2789(85)90011-9.

\bibitem{Rosenstein1993}
M.T. Rosenstein, J.J. Collins, C.J. De Luca,
A practical method for calculating largest Lyapunov exponents from small data sets,
\textit{Physica D: Nonlinear Phenomena} 65 (1--2) (1993) 117--134.
https://doi.org/10.1016/0167-2789(93)90009-P.

\bibitem{Parlitz2016}
U. Parlitz,
Estimating Lyapunov Exponents from Time Series,
in: C. Skokos, G.A. Gottwald, J. Laskar (Eds.), \textit{Chaos Detection and Predictability}, Lecture Notes in Physics, vol. 915, Springer, Berlin, Heidelberg, 2016, pp. 1--34.
https://doi.org/10.1007/978-3-662-48410-4\_1.

\bibitem{Pathak2017}
J. Pathak, Z. Lu, B.R. Hunt, M. Girvan, E. Ott,
Using machine learning to replicate chaotic attractors and calculate Lyapunov exponents from data,
\textit{Chaos} 27 (12) (2017) 121102.
https://doi.org/10.1063/1.5010300.

\bibitem{Ayers2022}
D. Ayers, J. Lau, J. Amezcua, A. Carrassi, V. Ojha,
Supervised machine learning to estimate instabilities in chaotic systems: Estimation of local Lyapunov exponents,
\textit{Quarterly Journal of the Royal Meteorological Society} 149 (2022) 1236--1262.
https://doi.org/10.1002/qj.4450.

\bibitem{Velichko2025}
A. Velichko, M. Belyaev, P. Boriskov,
A novel approach for estimating largest Lyapunov exponents in one-dimensional chaotic time series using machine learning,
\textit{Chaos: An Interdisciplinary Journal of Nonlinear Science} 35 (10) (2025) 101101.
https://doi.org/10.1063/5.0289352.

\bibitem{Mendes2016}
{\color{black}
E.M.A.M. Mendes, E.G. Nepomuceno,
A Very Simple Method to Calculate the (Positive) Largest Lyapunov Exponent Using Interval Extensions,
\textit{International Journal of Bifurcation and Chaos} 26 (13) (2016) 1650226.
https://doi.org/10.1142/S0218127416502266.
}

\bibitem{Velichko2026FEGPro}
{\color{black}
A. Velichko, N'Gbo N'Gbo, B. Carpentieri, M. Shams,
FEG-Pro: Forecast-Error Growth Profiling for Finite-Horizon Instability Analysis of Nonlinear Time Series,
\textit{arXiv:2605.17282 [nlin.CD]} (2026).
https://doi.org/10.48550/arXiv.2605.17282.
}

\bibitem{Velichko2026Negative}
{\color{black}
A. Velichko, N'Gbo N'Gbo, V.-T. Pham,
Equation-Free Period-Aware Forecast-Error Contraction for Estimating Negative Largest Lyapunov Exponents from Short Trajectory Ensembles,
\textit{arXiv:2608.05522 [nlin.CD]} (2026).
https://doi.org/10.48550/arXiv.2608.05522.
}

\bibitem{NGbo2026Mittag}
{\color{black}
N'Gbo N'Gbo, A. Velichko, B. Carpentieri,
Equation-Free Screening of Mittag--Leffler-Compatible Dynamics from Scalar Time Series via kNN Multi-Horizon Profiles,
\textit{arXiv:2607.08588 [math.DS]} (2026).
https://doi.org/10.48550/arXiv.2607.08588.
}

\bibitem{VelichkoHuyut2027}
{\color{black}
A. Velichko, M.T. Huyut,
State-specific respiratory signatures for affective and stress recognition: Interpretable respiratory markers, autocorrelation lags, and compact CNN models,
\textit{Biomedical Signal Processing and Control} 129, Part A (2027) 111373.
https://doi.org/10.1016/j.bspc.2026.111373.
}

\bibitem{Persohn2012}
K.J. Persohn, R.J. Povinelli,
Analyzing logistic map pseudorandom number generators for periodicity induced by finite precision floating-point representation,
\textit{Chaos, Solitons \& Fractals} 45 (3) (2012) 238--245.
https://doi.org/10.1016/j.chaos.2011.12.006.

\bibitem{Galias2021}
Z. Galias,
Periodic Orbits of the Logistic Map in Single and Double Precision Implementations,
\textit{IEEE Transactions on Circuits and Systems II: Express Briefs} 68 (2021) 3471--3475.
https://doi.org/10.1109/TCSII.2021.3081604.

\bibitem{Antonelli2018}
M. Antonelli, L. De Micco, H.A. Larrondo, O.A. Rosso,
Complexity of Simple, Switched and Skipped Chaotic Maps in Finite Precision,
\textit{Entropy} 20 (2) (2018) 135.
https://doi.org/10.3390/e20020135.

\bibitem{Kloewer2023}
M. Klöwer, P.V. Coveney, E.A. Paxton, T.N. Palmer,
Periodic orbits in chaotic systems simulated at low precision,
\textit{Scientific Reports} 13 (1) (2023) 11410.
https://doi.org/10.1038/s41598-023-37004-4.

\bibitem{Fan2021}
C. Fan, Q. Ding,
Analysis and resistance of dynamic degradation of digital chaos via functional graphs,
\textit{Nonlinear Dynamics} 103 (2021) 1081--1097.
https://doi.org/10.1007/s11071-020-06160-x.

\bibitem{Nepomuceno2017}
{\color{black}
E.G. Nepomuceno, S.A.M. Martins, G.F.V. Amaral, R. Riveret,
On the lower bound error for discrete maps using associative property,
\textit{Systems Science \& Control Engineering} 5 (1) (2017) 462--473.
https://doi.org/10.1080/21642583.2017.1387874.
}

\bibitem{Nepomuceno2018}
{\color{black}
E.G. Nepomuceno, S.A.M. Martins, B.C. Silva, G.F.V. Amaral, M. Perc,
Detecting unreliable computer simulations of recursive functions with interval extensions,
\textit{Applied Mathematics and Computation} 329 (2018) 408--419.
https://doi.org/10.1016/j.amc.2018.02.020.
}

\bibitem{Flajolet1990}
{\color{black}
P. Flajolet, A.M. Odlyzko,
Random Mapping Statistics,
in: J.-J. Quisquater, J. Vandewalle (Eds.), \textit{Advances in Cryptology -- EUROCRYPT '89}, Lecture Notes in Computer Science, vol. 434, Springer, 1990, pp. 329--354.
https://doi.org/10.1007/3-540-46885-4\_34.
}

\bibitem{Merah2021}
L. Merah, P. Lorenz, A.-P. Adda,
A New and Efficient Scheme for Improving the Digitized Chaotic Systems From Dynamical Degradation,
\textit{IEEE Access} 9 (2021) 88997--89008.
https://doi.org/10.1109/ACCESS.2021.3089913.

\bibitem{Li2024}
B. Li, K. Sun, H. Wang, W. Liu,
A delay-disturbance method to counteract the dynamical degradation of digital chaotic systems and its application,
\textit{Chaos, Solitons \& Fractals} 182 (2024) 114843.
https://doi.org/10.1016/j.chaos.2024.114843.

\bibitem{Wang2016}
Q. Wang, S. Yu, C. Li, J. Lü, X. Fang, C. Guyeux, J.M. Bahi,
Theoretical Design and FPGA-Based Implementation of Higher-Dimensional Digital Chaotic Systems,
\textit{IEEE Transactions on Circuits and Systems I: Regular Papers} 63 (3) (2016) 401--412.
https://doi.org/10.1109/TCSI.2016.2515398.

\bibitem{DeLaFraga2017}
L.G. de la Fraga, E. Torres-Pérez, E. Tlelo-Cuautle, C. Mancillas-López,
Hardware implementation of pseudo-random number generators based on chaotic maps,
\textit{Nonlinear Dynamics} 90 (2017) 1661--1670.
https://doi.org/10.1007/s11071-017-3755-z.

\bibitem{FloresVergara2019}
A. Flores-Vergara, E.E. García-Guerrero, E. Inzunza-González, O. López-Bonilla, E. Rodríguez-Orozco, J. Cárdenas-Valdez, E. Tlelo-Cuautle,
Implementing a chaotic cryptosystem in a 64-bit embedded system by using multiple-precision arithmetic,
\textit{Nonlinear Dynamics} 96 (2019) 497--516.
https://doi.org/10.1007/s11071-019-04802-3.

\bibitem{Murillo2022}
D. Murillo-Escobar, M. Murillo-Escobar, C. Cruz-Hernández, A. Arellano-Delgado, R. López-Gutiérrez,
Pseudorandom number generator based on novel 2D Hénon-Sine hyperchaotic map with microcontroller implementation,
\textit{Nonlinear Dynamics} 111 (2022) 6773--6789.
https://doi.org/10.1007/s11071-022-08101-2.

\bibitem{Huynh2019}
V.V. Huynh, A. Ouannas, X. Wang, V.-T. Pham, X.Q. Nguyen, F.E. Alsaadi,
Chaotic Map with No Fixed Points: Entropy, Implementation and Control,
\textit{Entropy} 21 (3) (2019) 279.
https://doi.org/10.3390/e21030279.

\bibitem{Nazare2026}
{\color{black}
T. Nazar\'e, M.S. Baptista, E.G. Nepomuceno,
Experimental Lower-Bound Error for Lyapunov Exponent Estimation Using Switch-Controlled Initial Conditions,
in: \textit{2026 IEEE International Symposium on Circuits and Systems (ISCAS)}, Shanghai, China, 24--28 May 2026.
https://doi.org/10.1109/ISCAS66217.2026.11562803.
}

\bibitem{Peixoto2018}
M.L.C. Peixoto, E.G. Nepomuceno, S.A.M. Martins, M.J. Lacerda,
Computation of the largest positive Lyapunov exponent using rounding mode and recursive least square algorithm,
\textit{Chaos, Solitons \& Fractals} 112 (2018) 36--43.
https://doi.org/10.1016/j.chaos.2018.04.032.

\bibitem{Zhou2019}
S. Zhou, X.-Y. Wang, Z. Wang, C. Zhang,
A novel method based on the pseudo-orbits to calculate the largest Lyapunov exponent from chaotic equations,
\textit{Chaos} 29 (2019) 033125.
https://doi.org/10.1063/1.5087512.

\bibitem{Tutueva2020}
{\color{black}
A.V. Tutueva, E.G. Nepomuceno, A.I. Karimov, V.S. Andreev, D.N. Butusov,
Adaptive chaotic maps and their application to pseudo-random numbers generation,
\textit{Chaos, Solitons \& Fractals} 133 (2020) 109615.
https://doi.org/10.1016/j.chaos.2020.109615.
}
\end{thebibliography}
\end{document}